\documentclass[12pt,fleqn]{article}
\usepackage{epsf}
\def\fnote#1#2{\begingroup\def\thefootnote{#1}\footnote{#2}\endgroup}
\makeatletter
\def\section{\@startsection {section}{1}{\z@}{3.5ex plus 1ex minus
    .2ex}{2.3ex plus .2ex}{\sc }}
\def\subsection{\@startsection{subsection}{2}{\z@}{3.25ex plus 1ex
minus
   .2ex}{1.5ex plus .2ex}{\small \sc }}
\def\appendix{\par\clearpage
  \setcounter{section}{0}
  \setcounter{subsection}{0}
  \@addtoreset{equation}{section}
  \def\@sectname{Appendix~}
  \def\theequation{\thesection.\arabic{equation}}
  \def\thesection{\Alph{section}}}
\makeatother
\makeatletter \@addtoreset{equation}{section} \makeatother
\renewcommand{\theequation}{\thesection.\arabic{equation}}

\def\ap#1#2#3{     {\it Ann. Phys. (NY) }{\bf #1} (19#2) #3}

\def\npb#1#2#3{    {\it Nucl. Phys. }{\bf B #1} (19#2) #3}
\def\plb#1#2#3{    {\it Phys. Lett. }{\bf B #1} (19#2) #3}

\def\zpc#1#2#3{    {\it Z. Phys. }{\bf C #1} (19#2) #3}

\def\nc#1#2#3{     {\it Nuovo Cim. }{\bf #1} (19#2) #3}

\def\ijmpa#1#2#3{  {\it Int. J. Mod. Phys. }{\bf A #1} (19#2) #3}

\def\eq#1{{eq.~(\ref{#1})}}

\let\vev\VEV
\def\abs#1{\left| #1\right|}
\def\mod#1{\abs{#1}}
\def\Im{\mathop{\mbox{Im}}}
\def\Re{\mathop{\mbox{Re}}}

\newcommand{\bea}{\begin{eqnarray}}
\newcommand{\beq}{\begin{equation}}
\newcommand{\eea}{\end{eqnarray}}
\newcommand{\eeq}{\end{equation}}

\newcommand{\spav}[1]{\parbox{1mm}{\vspace*{#1}}}

\renewcommand{\baselinestretch}{1.2}
\begin{document}
\begin{titlepage}
\begin{flushright}
{\tt SISSA 110/96/EP}
\end{flushright}
\vspace*{6cm}
\begin{center}
{\Large\bf A NEW ESTIMATE OF $\varepsilon '/\varepsilon$
\fnote{$\natural$}{Talk
given at the Workshop on K Physics, Orsay, France, May 30 - June 4, 1996.}
}\\
\spav{0.6cm}\\
{\large S. Bertolini} 
\\
{\em INFN, Sezione di Trieste, and}\\
{\em Scuola Internazionale Superiore di Studi Avanzati}\\
{\em via Beirut 4, I-34013 Trieste, Italy.}\\

\spav{7.0cm}\\
{\sc Abstract}
\end{center}
{
We discuss a new estimate 
of $\varepsilon '/\varepsilon$ in the kaon system. The present approach
is based on  the evaluation of the 
hadronic matrix elements of the  \mbox{$\Delta S =1$} 
effective quark lagrangian
by means of the chiral quark model, with the inclusion of
meson one-loop renormalization and NLO Wilson coefficients. 
The estimate here reviewed is fully consistent with the $\Delta I =1/2$
selection rule in $K\to \pi\pi$ decays, which is well reproduced within the
same framework.
}
\vfill

\end{titlepage}

\newpage
\setcounter{footnote}{0}
\setcounter{page}{2}

\section{Introduction}

The real part of $\varepsilon'/\varepsilon$ measures 
direct $CP$ violation in the decays of a neutral kaon in two 
pions. It is a fundamental quantity which has attracted a great
deal of theoretical as well as experimental work. Its
determination answers the question of whether
$CP$ violation is present only 
in the mass matrix of the neutral kaons (the superweak scenario) or is also
at work directly in the decays. 
 
On the experimental
front, the present results of CERN (NA31)
and Fermilab (E731)
are tantalizing insofar as the superweak scenario cannot be excluded and
the disagreement between the two outcomes leaves still a large uncertainty.

On the theoretical side, much has been accomplished, although the 
intrinsic difficulty
of a problem that encompasses scales as different as $m_t$ and $m_\pi$
weights against a decisive progress in the field.

The short distance (perturbative) QCD analysis has greatly profited
in recent years from the work done 
by the Munich~\cite{monaco}
and Rome~\cite{roma} groups who
computed the anomalous dimension matrix
of the ten relevant operators to the next-to-leading order (NLO), thus
reducing the uncertainty related to $\alpha_s$ at the 10\% level. 

The largest uncertainty arises in 
the long-distance part of the effective lagrangian, 
 the computation of which
 involves the evaluation of the
hadronic matrix elements of the quark operators. 
 
There exist two complete estimates of such hadronic matrix elements 
performed by the aforementioned
groups, and recently updated in ref.
\cite{martinelli} for the lattice (at least for some of
the operators). A recent review of $1/N_c$ results
is given in ref. \cite{buras}.

Both groups seem to agree on the difficulty of accommodating
within the standard model a value substantially
larger than $10^{-3}$ and obtain $10^{-4}$ as the preferred
scale for $\varepsilon'/\varepsilon$. 
This unexpectedly small value is the result of
the cancellation between gluonic and electroweak penguin 
operators~\cite{FR}. 
This cancellation, which depletes by an order of magnitude the natural 
magnitude of $\varepsilon'/\varepsilon$ in the SM, marres any theoretical
attempt of predicting direct CP violation with a precision better than
a factor two (due solely to the intrinsic 
short-distance uncertainty).

In view of that, it seems to us that independent estimates of
$\varepsilon'/\varepsilon$, even in phenomenological models, are   
desirable. In our opinion it is crucial
that a reliable evaluation of the hadronic matrix elements
first provides a consistent 
picture of kaon physics, starting
from the $CP$-conserving amplitudes and, in particular, by reproducing
the $\Delta I = 1/2$ selection rule, which governs most of these
amplitudes as well as the quantity 
$\varepsilon'/\varepsilon$ itself. 
We also feel that the same evaluation should pay particular
attention to the problem of achieving a satisfactory 
$\gamma_5$-scheme and scale independence in the matching between the
matrix elements and the Wilson coefficients, the absence of which 
undermines any estimate.

In ref.~\cite{ABEFL}, hereafter referred as I, we have completed the
study of the hadronic matrix elements of all the ten
operator of the $\Delta S =1$ effective quark
langrangian by means of the chiral quark model ($\chi$QM)~\cite{QM}
 and shown in~\cite{ABFL}, hereafter
referred as II, that the
inclusion of non-perturbative $O(\alpha_s N_c)$ corrections and the $SU(3)$
breaking effects in the
one-loop meson renormalization provided an
improved scale independence
and, more importantly, a good fit of the $\Delta I =1/2$ selection rule 
(see also the contribution of M. Fabbrichesi at this workshop).

In a recent paper \cite{BEF2} we presented a new computation
of $\varepsilon'/\varepsilon$ based on the $\chi$QM approach.
Our estimate takes advantage, as the existing ones, of
\begin{itemize}
\item
NLO results for the Wilson coefficients;
\item 
up-to-date analysis of the constraints on the mixing coefficient
Im $\lambda_t$.
\end{itemize}
Among the new elements introduced, the most relevant are 
\begin{itemize}
\item
A consistent evaluation of all hadronic matrix elements of the ten
effective quark operators in the $\chi$QM
(including next-to-leading $O(N)$ and  $O(\alpha_s N)$ contributions).
\item
Inclusion in the $\Delta S=1$ chiral lagrangian of the
complete bosonization $O(p^2)$ of the electroweak operators $Q_7$ and
$Q_8$. Some relevant $O(p^2)$ terms have been neglected
in all previous estimates;
\item
Inclusion of the meson-loop renormalization and scale dependence
of the matrix elements;
\item
Consistency with the $\Delta I = 1/2$ selection rule in kaon decays;
\item
Matching-scale and $\gamma_5$-scheme dependence of the results below the 
20\% level. 
\end{itemize}

Our prediction of $\varepsilon'/\varepsilon$
depends most sensitively on the value of 
the quark condensate,
the input parameter that controls penguin-diagram physics. 
For this reason, we  discuss a inclusive estimate based on a conservative
range of $\vev{\bar{q}q}$, as well as the variations of all the
other inputs: $m_t$,  $\Im \lambda_t$ (which depends, beside $m_t$ and
$m_c$, on
$\hat B_K$ and other mixing angles) and 
$\Lambda_{\rm QCD}$.
Such a procedure provides us with the range of
values for $\varepsilon'/\varepsilon$
that we consider to be the unbiased theoretical prediction of the standard model.
Unfortunately, this range turns out to be  rather large, spanning
from $-5 \times 10^{-3}$ to $1.4 \times 10^{-3}$. On the
other hand, it is as small as we can get without making some further
assumptions on the input parameters, assumptions that all the
other available estimates must make as well.

Such uncertainty notwithstanding, we agree in the end with 
refs.~\cite{martinelli,buras} that it is difficult to accommodate 
within the standard
model a value of $\varepsilon'/\varepsilon$ larger than $10^{-3}$.
As a matter of fact, our analysis points to definitely
smaller values, when not negative. This can be understood not so much as
a peculiar feature of the $\chi$QM prediction as the neglect in other estimates
of a class of contributions in the vacuum saturation approximation (VSA)
of the matrix elements of the leading electroweak operators, 
which enhance the destructive
interference between gluonic and electroweak penguins. This problem
is discussed in detail
in I and in ref. \cite{FL}. These new contributions
are responsible for the onset of the superweak regime for values of $m_t$
smaller than 200 GeV.
In our computation, it is the meson loop renormalization 
that in the end brings back  
$\varepsilon'/\varepsilon$ to positive values.

\section{Hadronic Matrix Elements and $B_i$ Factors}

The quark effective lagrangian at a scale $\mu < m_c$ can be written as
\beq
{\cal L}_{\Delta S = 1} = 
-\frac{G_F}{\sqrt{2}} V_{ud}\,V^*_{us} \sum_i \Bigl[
z_i(\mu) + \tau y_i(\mu) \Bigr] Q_i (\mu)  
\equiv -\frac{G_F}{\sqrt{2}} \sum_i C_i (\mu) Q_i (\mu)  
\, . 
\label{ham}
\eeq

The $Q_i$ are four-quark operators obtained by integrating out in the standard
model the vector bosons and the heavy quarks $t,\,b$ and $c$. A convenient
and by now standard
basis includes ten quark operators.
The functions $z_i(\mu)$ and $y_i(\mu)$ are the
 Wilson coefficients and $V_{ij}$ the
Koba\-ya\-shi-Mas\-kawa (KM) matrix elements; $\tau = - V_{td}
V_{ts}^{*}/V_{ud} 
V_{us}^{*}$. Following the usual parametrization of the KM matrix, in order to
determine $\varepsilon'/\varepsilon$, we only
need the $y_i(\mu)$, which control the $CP$-violating part of the amplitudes.

In paper I we have computed all hadronic matrix elements of the ten effective
quark operators in \eq{ham} in the framework of the $\chi$QM. 
The matrix elements of the first six operators were first computed in the 
$\chi$QM in ref. \cite{PdR}. The matrix 
elements are obtained by the integration of the constituent quarks by means of
dimensional regularization. The loop
integration leads to results that  depend on the scheme employed to 
deal with $\gamma_5$ but are scale independent. The renormalization-scale 
dependence
is introduced in our approach by the meson-loop renormalization of the
amplitudes, as explained in I. 
The meson-loop corrections together with the gluon-condensate 
contributions are the most relevant ingredients in reproducing
the $\Delta I = 1/2$ selection rule in $K\to \pi\pi$ decays, (as discussed
in II). 

The $\chi$QM results are expressed in a double power expansion on 
$M^2/\Lambda_{\chi}^2$ and $p^2/\Lambda_{\chi}^2$, 
where $M$ is a dimensionful parameter of the
model which  can be
interpreted as the constituent quark mass in mesons,  $p$
is a typical external momentum, and $\Lambda_{\chi}\simeq 1$ GeV is the chiral 
symmetry breaking scale. 

The value of $M$ is
constrained~\cite{B} by experimental data on the decay of $\pi^0$ and $\eta$ to
be
$
M = 223 \pm 9 \: \mbox{MeV}
$
($243 \pm 9$ MeV if higher order corrections are included).
Comparable values are found by using
vector-meson-dominance, or fitting all
input parameteres in the extended Nambu-Jona-Lasinio model~\cite{B2}.

The integration of the fermion degrees of freedom leads naturally
to an effective bosonic representation of the $\Delta S=1$ quark
lagrangian with the desired chiral properties. 
In I we have constructed the complete $O(p^2)$ chiral representation
of the lagrangian in \eq{ham}, where the local quark operators
$Q_i$ are represented by a linear combination of bosonic operators
$B_\alpha$, namely $Q_i \to \sum_\alpha G_\alpha(Q_i) B_\alpha$.
The effective quark lagrangian is therefore replaced
by 
\beq
{\cal L}^{\Delta S = 1}_\chi = 
-\frac{G_F}{\sqrt{2}}\ \sum_{i,\alpha}\ C_i (\mu)\ 
           G_\alpha (Q_i)\ B_\alpha  
\, . 
\label{chiham}
\eeq

As discussed at lenght in I, the chiral coefficients  
$G_\alpha$ determined 
via the $\chi$QM approach are 
$\gamma_5$-scheme dependent.
While the $\gamma_5$-scheme dependence arises in the $\chi$QM
from the integration of the chiral fermions, the explicit 
$\mu$-dependence is entirely due to the chiral loop
renormalization of the matrix elements:
\beq
\vev{b|{\cal L}^{\Delta S = 1}_\chi|a} = 
-\frac{G_F}{\sqrt{2}}\ \sum_{i,\alpha}\ C_i (\mu_{SD})\ 
           G_\alpha (Q_i)\ \vev{b|B_\alpha|a} (\mu_{LD}) 
\, , 
\label{chimatr}
\eeq
where we have labeled by $a$ and $b$ the initial and final bosonic states.
We remark that in our approach the $\mu$-dependence of the
chiral loops is not cancelled by higher order counterterms, as it
is usually required in the strong chiral lagrangian. 

The renormalization scale 
dependence is therefore determined order by order in the
energy expansion of the chiral lagrangian. 
In this respect there is no direct counterpart to the expansion
in strong and electromagnetic
couplings on which the short-distance
analysis is based and, accordingly, we refer to the
explicit $\mu$-dependence in the matrix elements as to
the long-distance (LD) or ``non-perturbative'' scale dependence.
A purely perturbative renormalization scale dependence is introduced
in the matrix elements by the NLO running of the quark condensate,
which we include whenever a comparison between values at
different scales is required. Otherwise, quark and gluon condensates
are considered in our approach as phenomenological parameters. 

Physical observables must not depend on the chosen scheme for handling
$\gamma_5$ in dimensional regularization and the value of
the renormalization scale $\mu$. 
Our aim is to test whether the estimate of observables
is consistently improved by matching the ``long-distance'' 
$\gamma_5$-scheme and $\mu$ dependences so obtained
with those present in the short-distance analysis (in particular
we identify $\mu_{SD}$ with $\mu_{LD}$).
Whether and to what extent
such an improvement is reproduced for many observables and
for a consistent set of parameters, might tell us how well low-energy
QCD is modelled in the $\chi$QM-chiral lagrangian approach  that
we have devised.

In II, we have shown that minimizing the $\gamma_5$-scheme dependence
of the physical 
isospin $I=0$ and 2 amplitudes determines a range for the parameter
$M$  between 160 and 220 MeV. 
In II, it  was also found that the $\mu$ dependence induced
by the Wilson coefficients is substantially reduced by that of the
hadronic matrix elements. 

These issues become crucial for $\varepsilon'/\varepsilon$ where
the $\gamma_5$-scheme dependence induced by the Wilson coefficients determines
an uncertainty as large as 80\% when using
the $1/N_c$ hadronic matrix elements (see for instance ref.~\cite{monaco})
which are scheme independent.

In order to test the $\mu$ independence
of $\varepsilon'/\varepsilon$ we vary the matching scale between 0.8 and 1.0
GeV, the highest energy up to which we trust the chiral loop corrections computed 
in I.
We find that, in spite of the fact that some of the Wilson coefficients
vary in this range by up to 50\%, the matching with our matrix elements
reduces the $\mu$-dependence in $\varepsilon'/\varepsilon$ below
20\% in most of the parameter space. We consider this improved stability
a success of the approach.

In order to discuss our results for the hadronic 
matrix elements it is convenient to introduce the effective factors
$
B_i^{(0,2)} \equiv \langle Q_i \rangle _{0,2}^{\chi {\rm QM}} /
                   \langle Q_i \rangle _{0,2}^{\rm VSA} \, ,
$
which give the ratios between our hadronic matrix elements and those of the
VSA. They are a useful way of comparing different evaluations.

In table 1 we collect the $B_i$ factors for the ten
operators in the isospin 0 and 2 channels. 
The values of the $B_i$ depend on the
scale at which the matrix elements
 are evaluated,  the input parameters and
$M$; moreover, in the $\chi$QM they depend on the
$\gamma_5$-scheme employed.
We have given in table 3 a representative 
example of their values and variations.
\renewcommand{\baselinestretch}{1}
\begin{table}[t]
\begin{small}
\begin{center}
\begin{tabular}{|c||c|c|c|c|}
\hline
 & \multicolumn{2}{|c|}{\rm HV} & \multicolumn{2}{c|}{\rm NDR} \\ 
\cline{2-5}
\cline{2-5}
 & $\mu = 0.8$ GeV & $\mu = 1.0$ GeV & $\mu = 0.8$ GeV & $\mu = 1.0$ GeV\\
\hline
$B^{(0)}_1$  & 10.6 & 11.1 & 10.6 & 11.1\\
\hline
$B^{(0)}_2$  & 2.8 & 3.0  & 2.8 & 3.0\\
\hline
$B^{(2)}_1$  & 0.52 & 0.55 &  0.52 & 0.55 \\
\hline
$B^{(2)}_2$ & 0.52 & 0.55 & 0.52 & 0.55 \\
\hline
$B_3$ & $-2.9$ & $-3.0$ & $-3.7$ & $-3.9$\\
\hline
$B_4$ & 1.8 & 1.9 & 1.0 & 1.1\\
\hline
$B_5 = B_6$ & $1.7 \div 0.61$ & $1.8 \div 0.64$ & $1.0 \div 0.38$ & $1.1 \div 0.40$\\
\hline
$B_7^{(0)}$ &$3.0 \div 2.2$ &  $3.3 \div 2.4$ & $2.9 \div 2.2$  & $3.2 \div 2.3$ \\
\hline
$B_8^{(0)}$ &$3.3 \div 2.2$& $3.6 \div 2.4$ &  $3.2 \div 2.2$ & $3.5 \div 2.4$ \\
\hline
$B_9^{(0)}$ &3.9 & 4.0 & 3.5  & 3.6\\
\hline
$B_{10}^{(0)}$ &4.4 & 4.7 & 5.6 & 5.9\\
\hline
$B_7^{(2)}$ &$2.7 \div 1.5$ & $3.0 \div 1.5$ & $2.7 \div 1.4$ & $2.9 \div 1.5$\\
\hline
$B_8^{(2)}$ &$2.1 \div 1.4 $& $2.3 \div 1.5$ & $2.1 \div 1.4$ & $2.3 \div 1.5$\\
\hline
$B_9^{(2)}$ &0.52 & 0.55 & 0.52 & 0.55 \\
\hline
$B_{10}^{(2)}$ & 0.52 & 0.55 & 0.52 & 0.55 \\
\hline
\end{tabular}
\end{center}
\end{small}
\caption{$B_i$ factors in the $\chi$QM (including meson-loop
renormalizations) at two different scales: $\mu$ = 0.8 and 1.0 GeV and in the
two $\gamma_5$-schemes. The results shown are given for $M = 220$ MeV
and
$\vev{\alpha_s GG}/\pi = (376$ MeV$)^4$, while the range given
for $B_{5-8}$
corresponds to varying the quark condensate between $(-200$ MeV$)^3$
and $(-280$ MeV$)^3$.}
\end{table}

The values of $B_1^{(0)}$ and $B_2^{(0)}$ show 
that the corresponding
hadronic matrix elements in the $\chi$QM are, once non-factorizable
contributions and meson renormalization have been included, respectively
about ten and three times
larger than their VSA values. 
At the same time, 
$B_1^{(2)}$ and $B_2^{(2)}$ turn out to be 
at most half of what found in the VSA. 
These features make
it possible for the selection rule to be reproduced in the $\chi$QM, as extensively
discussed in II.

For comparison, in the $1/N_c$ approach of ref. \cite{1/N}, the inclusion of
meson-loop renormalization through a cutoff regularization, leads, at the
scale of 1 GeV, to $B_1^{(0)} = 5.2$, $B_2^{(0)} = 2.2$ and 
$B_1^{(2)} = B_2^{(2)} = 0.55$, a result that is not sufficient to
reproduce the $\Delta I = 1/2$ rule.
The similarity of the values $B_1^{(2)} = B_2^{(2)} = 0.55$ obtained in
the $\chi$QM with the corresponding $1/N_c$ results 
is remarkable, and yet
a numerical coincidence, since the suppression originates from gluon
condensate corrections in the $\chi$QM, whereas it is the effect of
the meson loop renormalization (regularized via explicit cut-off)
in the analysis of ref. \cite{1/N}. 

The values of the penguin matrix elements $\vev{Q_3}$ 
and $\vev{Q_4}$ in the $\chi$QM lead to rather large $B_i$ factors. 
In the case of $Q_3$, the $\chi$QM result has the opposite sign of the VSA
result and $B_3$ is negative.
This is the effect of the large non-perturbative gluon
correction. 

Regarding the gluon penguin operator $Q_6$ (and $Q_5$), we find that 
the $\chi$QM gives
a result consistent with the VSA (and the $1/N_c$ approach), 
$B_6$ ($B_5$) being approximately equal to two for
small values of the quark condensate and one-half at larger values. It is the
quadratic dependence (to be contrasted to the linear dependence in the
$\chi$QM) of the VSA matrix element for the penguin operators that
it responsible for the different weight of these operators at different values
of the quark condensate. 
The lattice estimate at $\mu = 2$ GeV 
for these operators 
gives $B_5 = B_6 = 1.0 \pm 0.2$~\cite{martinelli}.

The electroweak $B_i$ factors are all larger in the $\chi$QM than in the VSA,
except for $B_{9,10}^{(2)}$ that are about
 1/2 in the HV and in the NDR schemes. 
The lattice estimate at $\mu = 2$ GeV
in this case yields $B_{7,8}^{(2)} = 1.0 \pm 0.2$ and
$B_9^{(2)} = 0.62 \pm 0.10$~\cite{martinelli}.
For new preliminary lattice results
see the contribution of G. Kilcup at this workshop.

The most relevant result for
$\varepsilon'/\varepsilon$ is the value of $B_8^{(2)}$ which ranges
from 1.5 to 2 times that of the VSA. This increase is due to two independent
reasons. On the one hand, we found two new terms 
in the chiral lagrangian that
have not been included so far in the VSA estimate of the $Q_{7,8}$
matrix elements. 
The chiral coefficients of these terms are 
computed in the $\chi$QM approach---as
discussed in detail in I. 

From this point of view,
what we have referred to as VSA---and used in table 1 as normalization for the 
$Q_{7,8}$ operators---is not the complete VSA result. 
The inclusion of the new terms amounts up to a 60\%
increase of $B_{7,8}^{(2)}$ for small values of $\vev{\bar qq}$ 
in the chosen range and down to about 10\%
for large values; smaller effects are found in the case of  $B_{7,8}^{(0)}$. 
On the other hand, the meson-loop renormalization 
associated with the new chiral terms is large (see
I) and adds up to reproduce the results shown in table 1.
The increase in importance of the operator $Q_8$ with
respect to $Q_6$ turns into a
 more effective cancellation between the two operators for large
values of the quark condensate while at smaller
values the gluon penguin contribution prevails.

\section{Anatomy of $\varepsilon'/\varepsilon$ in the $\chi$QM}

The quantity
 $\varepsilon'/\varepsilon$ can be written as
\beq
\frac{\varepsilon '}{\varepsilon} =  
\frac{G_F \omega}{2\mod{\epsilon}\Re{A_0}} \:
\mbox{Im}\, \lambda_t \: \:
 \left[ \Pi_0 - \frac{1}{\omega} \: \Pi_2 \right] \, ,
\label{eps}
 \eeq
where, referring to the $\Delta S=1$ quark lagrangian of \eq{ham},
\beq
 \Pi_0 =   \sum_i y_i \, \langle  Q_i  \rangle _0 \ ,\quad\quad
 \Pi_2 =   \sum_i y_i \, \langle Q_i \rangle_2  + 
\omega \: \sum_i y_i \, \langle  Q_i  \rangle _0  \: \Omega_{\eta +\eta'} \ ,
\eeq
and
$
\mbox{Im}\, \lambda_t \equiv \Im V_{td}V^*_{ts} \ .
$

The quantity $\Omega_{\eta +\eta'}$ includes the effect of the isospin-breaking 
mixing between $\pi^0$ and the etas.

A range for Im $\lambda_t$ is determined from the experimental value of
$\varepsilon$ as a function of $m_t$ and the other
relevant parameters involved in the theoretical estimate.
We will use the most recent NLO results for the QCD correction factors
$\eta_{1,2,3}$ (see the contribution of U. Nierste at this workshop)
and vary the $\Delta S = 2$ hadronic parameter $\hat B_K$ 
in the rather conservative range
 $
 \hat B_K = 0.55 \pm 0.25 \ ,
 $
 that 
 encompasses both the $\chi$QM model prediction~\cite{BK} and other current
 determinations~\cite{PP}.

\renewcommand{\baselinestretch}{1}
\begin{figure}[t]
\epsfxsize=10cm
\centerline{\epsfbox{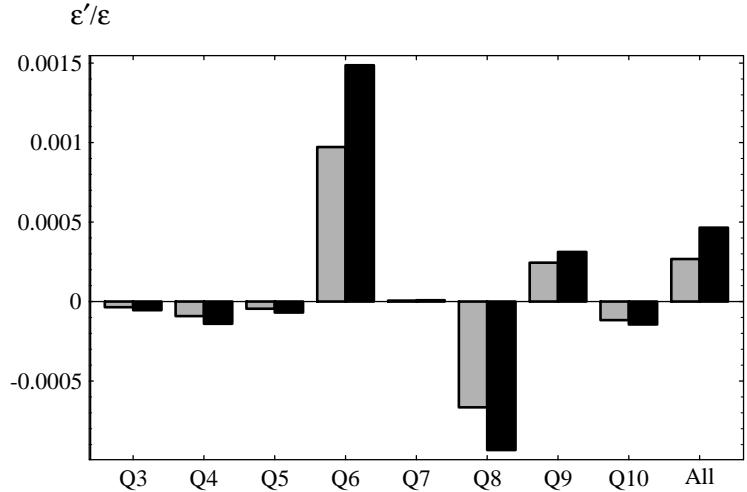}}
\caption{Histograms of the partial contributions to $\varepsilon'/\varepsilon$
of the height relevant operators for 
$\vev{\bar{q}q} (0.8\ {\rm GeV}) = (-200$ MeV)$^3$,
 $m_t^{\rm pole} = 180$ GeV,
$\Im \lambda_t = 1.3 \times 10^{-4}$ 
and $\Lambda_{\rm QCD}^{(4)} = 350$ MeV.
Gray (black) histograms represent the contribution of each operator 
without (with)
meson-loop renormalization. The last two histograms correspond to the sum of
all contributions.}
\end{figure}
 This procedure gives two possible ranges
for $\Im \lambda_t \simeq \eta |V_{us}| |V_{cb}|^2$, 
which correspond to having the KM phase in
the I or II quadrant ($\rho$ positive or negative, respectively). 
For example, for $m_t^{\rm pole} = 180$ GeV 
($\overline{m}_t(m_W) \simeq 183$ GeV) and 
$\Lambda_{\rm QCD}^{(4)}= 350$ MeV we find
 \beq
 1.1 \times 10^{-4} \leq \Im \lambda_t \leq 1.9 \times 10^{-4}
 \ \ \ {\rm and} \ \ \
 0.75 \times 10^{-4} \leq \Im \lambda_t \leq 1.9 \times 10^{-4}
 \eeq 
 in the first and second quadrant respectively. For the range of $\hat B_K$ 
considered, varying all the other
parameters (including $m_t$ and $\Lambda_{\rm QCD}$) 
affects the above limits on $\Im \lambda_t$ by less than 20\%. 
In particular, the
upper bound on $\Im \lambda_t$ is stable, 
becoming a sensitive function of the input parameters
only if we consider $\hat B_K > 0.5$.
In other words, we agree with ref.~\cite{PP} that
it is the theoretical uncertainty on the hadronic  
$\Delta S=2$ matrix element 
that controls the uncertainty on the determination
of $\Im \lambda_t$.

It is useful to consider the individual contribution
to $\varepsilon'/\varepsilon$
 of each
 of the quark operators. We have depicted them as histograms, where 
 the grey (black) one stands for the
 contribution before (after) meson-loop renormalization. Henceforth all
results are given for $M = 220$ MeV in the HV scheme.

\begin{figure}
\epsfxsize=10cm
\centerline{\epsfbox{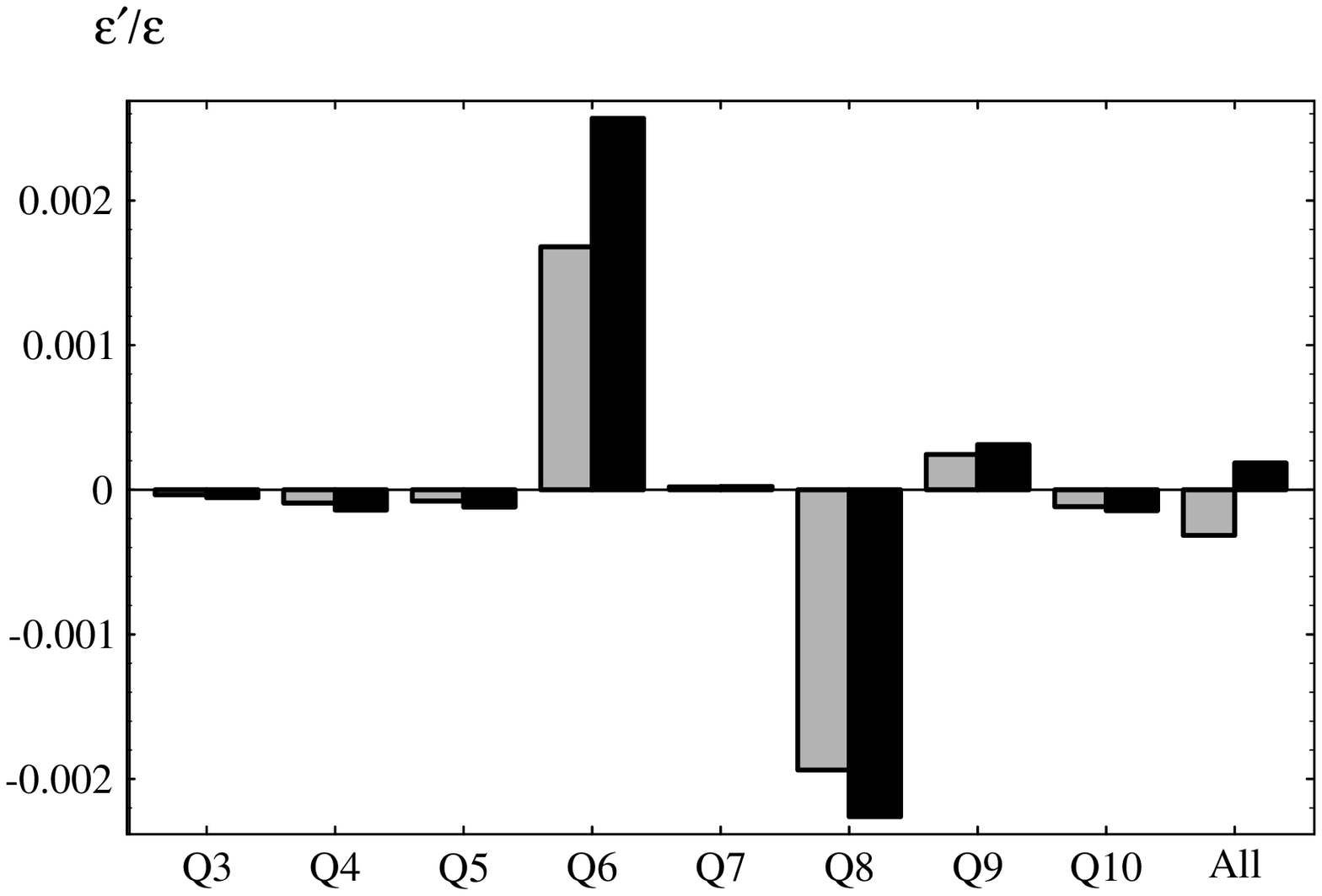}}
\caption{Same as Fig. 1 for 
$\vev{\bar{q}q} (0.8\ {\rm GeV}) = (-240$ MeV)$^3$.}
\end{figure}
It is clear from the 
 histograms of Fig. 1, 2 and 3 that the two 
 dominating operators are $Q_6$ and $Q_8$. Yet, since they give contributions
 approximately of the same size and opposite in sign, the final value turns out
 to be relatively small and of size comparable to that of most of
 the other operators. This result is at the origin the large theoretical
uncertainty as well as the unexpected smallness of  
$\varepsilon'/\varepsilon$.
 
The same histograms serve the purpose of showing that the meson-loop 
 renormalizations  are
 crucial not only in the overall size of each contribution but also
in determining the sign of the final result (see Fig. 2). 
These corrections are here consistently
included in the estimate for the first time.
  
The role of the operator $Q_4$ turns out to be
 marginal in our approach. In comparing 
this result with
that of the $1/N_c$ framework~\cite{buras} (see also
the final tables in ref. \cite{BEF} where we reproduce the individual $1/N_c$
contributions for the standard ten operators), it
should be recalled that in the above analysis
the $Q_4$ operator is written in terms of $Q_1$, $Q_2$ 
and $Q_3$ and that its values is therefore influenced by the $B_i$ factors
assigned to the former matrix elements. 
In particular, while $B_1$ and $B_2$ are  in ref.~\cite{buras} 
requested to be large
in order to account for the $\Delta I =1/2$ rule, $B_3$ 
is assigned the
value of 1. Such a procedure produces a rather large value for the
matrix element of $Q_4$. In our approach, we see that in fact also $B_3$ is 
large (and negative!) and that $Q_4$, 
once written in terms of the other operators,
is small, as found in the direct estimate.

\begin{figure}
\epsfxsize=10cm
\centerline{\epsfbox{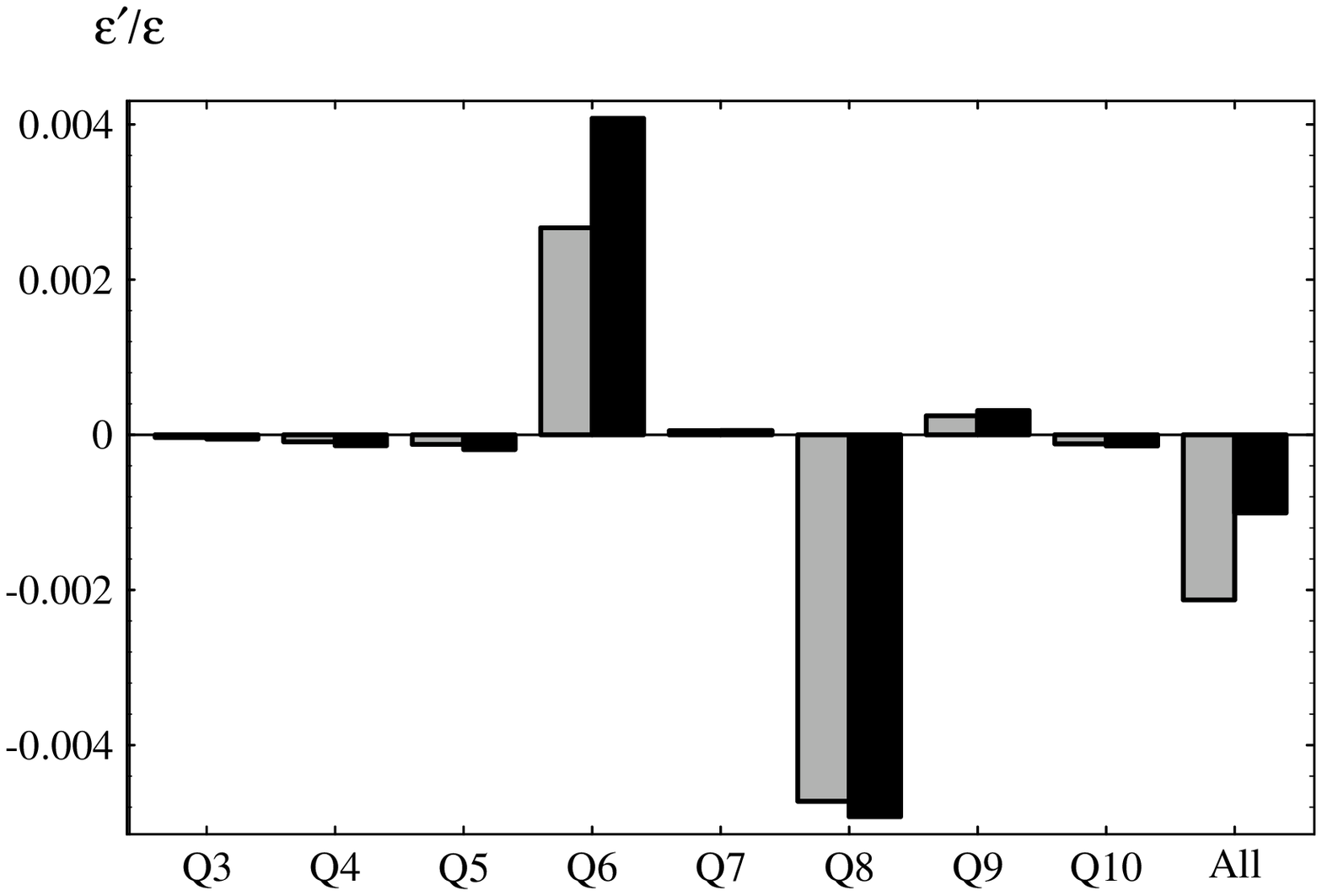}}
\caption{Same as Fig. 1 for 
$\vev{\bar{q}q} (0.8\ {\rm GeV}) = (-280$ MeV)$^3$.}
\end{figure}
We hope that this discussion has convinced the reader that
the quantity $\varepsilon'/\varepsilon$ is difficult to estimate even with 
a factor two uncertainty. 
We think that
only the order of magnitude can be predicted in a reliable manner.
The reason is very simple: the final value is the result of the cancellation
between two, approximately equal in size, contributions. Accordingly, any
uncertainty is most likely amplified by an order of magnitude and we are 
unfortunately  dealing with
rather large ones.
It is also important to understand that these considerations 
hold for any theoretical approach.

Keeping this in mind, by varying all parameters in the allowed ranges and, in
particular, 
taking the quark condensate, which is the major source of
uncertainty, between $(-200\ {\rm MeV})^3$ and $(-280\ 
{\rm MeV})^3$
we find
\beq
-27\times 10^{-4}\ < \varepsilon '/\varepsilon < \ 9 \times 10^{-4}\, ,
\eeq
where we have kept $\Lambda_{\rm QCD} ^{(4)}$ fixed at its central value. A larger
range, 
\beq
-50\times 10^{-4}\ < \varepsilon '/\varepsilon < \ 14 \times 10^{-4}\, ,
\label{grande}
\eeq
is obtained by varying $\Lambda_{\rm QCD}^{(4)}$  as well.

It should be stressed that the large range of negative values that we obtain
is a consequence of two characteristic features of the matrix elements:
i) the enhancement of the size of the electroweak matrix elements 
$\vev{Q_{8,7}}$ 
due to the coherent effects of the 
additional $O(p^2)$ contributions so far neglected 
(see discussion in sect. 5) and the chiral loop corrections; 
ii) the linear dependence 
on $\vev{\bar q q}$ of the leading gluon penguin matrix elements
compared to the quadratic dependence of the leading terms in the electroweak
matrix elements, which makes the latter prevail for large values
of the quark condensate. 
The effect of i) represents 
an enhancement of the leading electroweak matrix elements
up to a factor two with respect to
the vacuum insertion approximation and present $1/N_c$ estimates 
(see table 3), while feature ii) is absent in the $1/N_c$ approach,
the quark condensate dependence being always quadratic.  

To provide a somewhat more restrictive estimate 
we may assume for the quark condensate the QCD Sum Rules improved
PCAC result~\cite{narison}, namely 
$\vev{\bar qq} = -(221\: \pm 17\ {\rm MeV})^3$ at our matching
scale $\mu = 0.8$ GeV,
 and thus find
\beq
\varepsilon '/\varepsilon  = \left\{ \begin{array}{ll}
 4.5 \: ^{+4.1}_{-5.4} \,\times \,10^{-4} & {\rm quadrant \: I} \\
 3.9 \: ^{+5.0}_{-4.5} \,\times \,10^{-4} & {\rm quadrant \: II} \, .
 \end{array} \right. 
 \label{best} 
\eeq
whose average gives  
$\varepsilon '/\varepsilon = ( 4 \pm 5) \times 10^{-4} $.

The range of the quark condensate on which the above
 estimate is based, is not favorite  
 by our analysis of the
 $\Delta I = 1/2$ selection 
rule in the $\chi$QM. Higher values accommodate more naturally the rule, 
at least for a constituent mass $M \simeq
 220$ MeV---the value at which we also find $\gamma_5$-scheme independence of
 $\varepsilon'/\varepsilon$.
For values of the quark condensate in the range of Figs. 2 and 3
the central value of
$\varepsilon'/\varepsilon$ shifts toward the superweak regime,
and the role of meson loop corrections becomes crucial.
For instance, by taking 
the simple PCAC result for the strange quark condensate, 
$\vev{\bar qq} = -(244\: \pm 9\ {\rm MeV})^3$ at $\mu = 0.8$ GeV,
we find
\beq
\varepsilon '/\varepsilon  = \left\{ \begin{array}{ll}
 1.4 \: ^{+6.5}_{-5.5} \,\times \,10^{-4} & {\rm quadrant \: I} \\
 1.2 \: ^{+9.3}_{-4.0} \,\times \,10^{-4} & {\rm quadrant \: II} \, .
 \end{array} \right.
 \label{secondbest} 
\eeq
Actually, for such a range of $\vev{\bar q q}$, negative 
central values of $\varepsilon '/\varepsilon$ in both quadrants
are obtained due to the extra terms of
the bosonization of the electroweak operators $Q_7$ and $Q_8$ 
neglected in the previous estimates. Only after the inclusion of the
meson-loop renormalization $\varepsilon '/\varepsilon$ turns  
to the positive central values of \eq{secondbest}.

To have an overview of the present theoretical status for $\varepsilon'/\varepsilon$
in Fig. 4 we have summarized the predictions of three theoretical approaches
and compared them with the present 1 $\sigma$ 
experimental results (labeled bands). 
\renewcommand{\baselinestretch}{1}
\begin{figure}[thb]
\epsfxsize=10cm
\centerline{\epsfbox{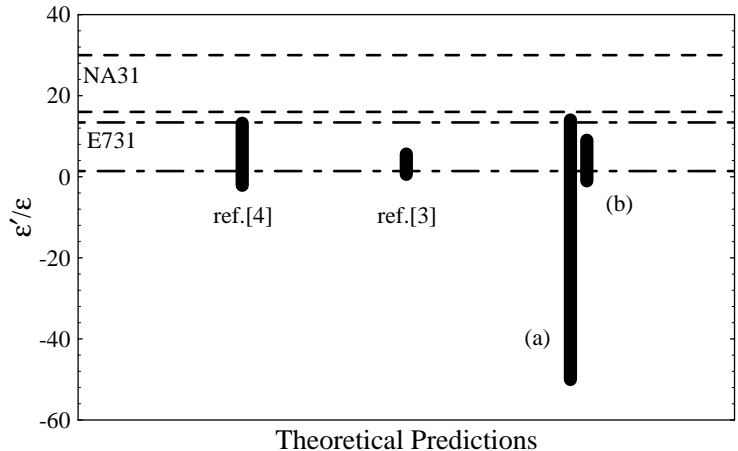}}
\caption{Present status of theoretical predictions and experimental values
for $\varepsilon'/\varepsilon$ (in units of $10^{-4}$). 
The most recent $1/N_c$ [4] and lattice [3] estimates are compared
to (a) our unbiased estimate (3.5), 
(b) our more restrictive estimate (3.6).}
\end{figure}

The smaller error in the lattice estimate 
originates in the Gaussian treatment of
the uncertainty in the input parameters 
with respect to the flat 1$\sigma$ error
included in the other estimates.

\section{Outlook}

Our phenomenological analysis, 
based on the simplest implementation
of the $\chi$QM  and chiral lagrangian methods,
takes advantage of the observation that the $\Delta I = 1/2$ selection
rule in kaon decays is well reproduced in terms of three basic 
parameters (the constituent quark mass $M$ and the quark and gluon 
condensates) in terms of which all hadronic matrix elements
of the $\Delta S=1$ lagrangian can be expressed.

We have used the best fit of the selection rule to constrain the allowed
ranges of $M$, $\vev{\bar q q}$ and $\vev{GG}$ and we have fed them
in the analysis of $\varepsilon '/\varepsilon$. 

Nonetheless,
the error bars on the prediction of $\varepsilon'/\varepsilon$
remain large.
This is due to two conspiring features: 1) the destructive interference
between the large hadronic matrix elements of $Q_6$ and $Q_8$ which 
enhances up to an order of magnitude any related uncertainty in the final 
prediction (this feature is general and does not depend on the
specific approach); 2) the fact that large quark-condensate values
are preferred in fitting the isospin zero $K^0\to \pi\pi$ amplitude at $O(p^2)$
(which is a model dependent result).

Whereas little can be done concerning point 1) which makes 
difficult any theoretical attempt to predict $\varepsilon'/\varepsilon$ 
with a precision better than a factor two, an improvement on 2) can be 
pursued within the present approach. 

Two lines of research are in progress.
On the one hand, we are extending the analysis to $O(p^4)$ in the 
chiral expansion to gain  better
precision on the hadronic matrix elements and to determine in a 
self-consistent way the polinomial contributions from the chiral loops;
preliminary results indicate that the $\Delta I = 1/2$ rule is 
reproduced for smaller values of the gluon and quark condensates, thus reducing
our error bar, in the direction shown by our more restrictive estimate.
On the other hand, we are studying the $\Delta S = 2$ sector
to determine at the same order of accuracy
$\hat{B}_K$ and the $K_L$--$K_S$ mass difference by
including in the latter the interference with long-distance contributions 
that can be self-consistently computed in the present approach.

Whether this program is successfull may better determine how much of the 
long range dynamics of QCD is embedded in the present approach
and increase our confidence on the predictions of unknown observables.

%
%

\renewcommand{\baselinestretch}{1}


\begin{thebibliography}{99}

{\small



\bibitem{monaco}  A.J. Buras, M. Jamin and M.E. Lautenbacher,
\npb{408}{93}{209}.

\bibitem{roma} M. Ciuchini, E. Franco, G. Martinelli and L. Reina,
\npb{415}{94}{403}; \plb{301}{93}{263}.

\bibitem{martinelli} M. Ciuchini, E. Franco, G. Martinelli and L. Reina,
in {\it The Second
DA$\Phi$NE Physics Handbook}, eds. L. Maiani et al. (Frascati, 1995);
\zpc{68}{95}{239} and references therein.

\bibitem{buras} G. Buchalla, A.J. Buras and M.E. Lautenbacher, 
{\em Weak Decays beyond Leading Logarithms}, hep-ph/95112380, to appear
in {\em Rev. Mod. Phys.}

\bibitem{FR} J. Flynn and L. Randall, \plb{224}{89}{221}; 
Erratum, \plb{235}{90}{412};
M. Lusignoli, \npb{325}{89}33;
G. Buchalla, A.J. Buras and M.K. Harlander, \npb{337}{90}{313}.

\bibitem{ABEFL} V. Antonelli, S. Bertolini, J.O. Eeg,
M. Fabbrichesi and E.I. Lashin, \npb{469}{96}{143}.

\bibitem{QM} K. Nishijima, \nc{11}{59}{698};
F. Gursey, \nc{16}{60}{230} and \ap{12}{61}{91}; 
J.A. Cronin, {\it Phys. Rev.} {\bf 161} (1967) 1483; 
S. Weinberg, {\it Physica} {\bf  96A} (1979) 327; 
A. Manohar and H. Georgi, \npb{234}{84}{189}; 
A. Manohar and G. Moore, \npb{243}{84}{55}.

\bibitem{ABFL}  V. Antonelli, S. Bertolini, 
M. Fabbrichesi and E.I. Lashin, \npb{469}{96}{181}. 

\bibitem{BEF2}  S. Bertolini, J.O. Eeg and  
M. Fabbrichesi, SISSA 103/95/EP, to appear in {\it Nucl. Phys.} {\bf B} (1996). 

\bibitem{FL} M. Fabbrichesi and E.I. Lashin, preprint SISSA 74/96/EP.


\bibitem{PdR}  A. Pich and E. de Rafael, \npb{358}{91}{311}.

\bibitem{B} J. Bijnens, \ijmpa{8}{93}{3045}.

\bibitem{B2} J. Bijnens, Ch. Bruno and
E. de Rafael, \npb{390}{93}{501}.

\bibitem{1/N} W.A. Bardeen A.J. Buras and J.-M. G\'{e}rard, 
\plb{192}{87} 138; 
W.A. Bardeen, A.J. Buras and J.-M. G\'{e}rard, \npb{293}{87}{787}.


\bibitem{BK} 
Ch. Bruno, \plb{320}{94}{135}; 
V. Antonelli, S. Bertolini, M. Fabbrichesi and E.I. Lashin,
preprint SISSA 20/96/EP.

\bibitem{PP} A. Pich and J. Prades, \plb{346}{95}{342}.

\bibitem{BEF}  S. Bertolini, J.O. Eeg and M. Fabbrichesi, \npb{449}{95}{197}.

\bibitem{narison} S. Narison, hep-ph/9504333.
}

\end{thebibliography}
\end{document}